\documentclass[pra,twocolumn,superscriptaddress,showpacs,floatfix]{revtex4}
\usepackage{amsfonts}

\usepackage{epsfig,amsmath,amssymb,latexsym}

\usepackage{subfigure}

\begin{document}

\title{ Local Kondo entanglement and its breakdown in an effective two-impurity Kondo model }

\preprint{1}

\author{Yuxiang Li}
  \affiliation{Condensed Matter Group,
  Department of Physics, Hangzhou Normal University, Hangzhou 310036, China}

\author{Xiao-Yong Feng}
  \affiliation{Condensed Matter Group,
  Department of Physics, Hangzhou Normal University, Hangzhou 310036, China}
\author{Jianhui Dai}
  \affiliation{Condensed Matter Group,
  Department of Physics, Hangzhou Normal University, Hangzhou 310036, China}


\begin{abstract}
Competition between the Kondo effect and
Ruderman-Kittel-Kasuya-Yosida interaction in the two-impurity Kondo
problem can be phenomenologically described by the Rasul-Schlottmann
spin model. We revisit this model from the quantum entanglement
perspective by calculating both the inter-impurity entanglement and
the local Kondo entanglement, the latter being the entanglement
between a local magnetic impurity and its spatially nearby
conduction electron. A groundstate phase diagram is derived and a
discontinuous breakdown of the local Kondo entanglement is found at
the singular point, associated concomitantly with a jump in the
inter-impurity entanglement. An entanglement monogamy holds in the
whole phase diagram. Our results identify the important role of the
frustrated cross-coupling and demonstrate the local characteristic
of the quantum phase transition in the two-impurity Kondo problem.
The implications of these results for Kondo lattices and quantum
information processing are also briefly discussed.
\end{abstract}

 \maketitle

\section{Introduction}

The past decade has witnessed the growing power of quantum
entanglements developed from the quantum information in
understanding the novel quantum states and quantum phase transitions
in solids\cite{Osterloh,Osborne,Vidalprl,Amico,Horodeckirmp}.
Condensed matter systems involve spin-$\frac{1}{2}$ objects as
natural qubits and various spin exchange interactions as sources of
quantum correlations. The present paper will explore from the
quantum entanglement perspective the variable Kondo effect in a
two-impurity Kondo model (TIKM), an intriguing problem involving two
typical kinds of spin exchanges.

It is known that Kondo systems, consisting of both itinerant
electrons and local magnetic moments, naturally involves the
single-ion Kondo effect and the Ruderman-Kittel-Kasuya-Yosida (RKKY)
interaction. The competition between them plays a key role in
correlated systems ranging from diluted mangetic alloys to heavy
fermion compounds\cite{Hewson,Coleman1}. The issue has been
intensively investigated within the TIKM\cite{JKW81} where, in
addition to the antiferromagnetic (AFM) Kondo coupling between a
magnetic impurity (or local moment) and its spatially nearby
conduction electron, the two magnetic impurities are also coupled
due to the RKKY interaction. There are two stable fixed points: the
strong Kondo coupling limit with each impurity spins being
completely quenched by the Kondo effect, and the strong RKKY
interaction limit with a different antiferromagnetic spin
singlet\cite{JKW81,Jones88}. The later situation has usually an AFM
ordered groundstate in the concentrated Kondo lattice
case\cite{Doniach77}. The TIKM has been realized in nanoscale
devices where the observed Kondo signature varies with tunable RKKY
interaction\cite{Glazman,Craig04}. In addition to the conventional
spin-density wave quantum phase transition, the variation of Kondo
effect and its competition with magnetic order in the Kondo lattice
systems may result in a local type of quantum phase transition
signaled by a breakdown of Kondo effect.\cite{Si1}

Theoretically, the primary focus is on the intermediate regime where
the single-ion Kondo and RKKY energy scales, represented by $T_K$
and $J_R$ respectively, are comparable to each other so that the
physical properties of the two stable fixed points
crossover\cite{Jones88,Jones89}. Of particular interesting is the
case when the crossover is sharpened leading to a phase transition.

Early numerical renormalization group\cite{Jones88} and conformal
field theory studies\cite{Affleck1} on the particle-hole symmetric
TIKM revealed an unstable interacting fixed point at a finite ratio
$J_R/T_K\approx 2.2$ where non-Fermi liquid behaviors such as the
divergent staggered susceptibility and specific heat coefficient
were observed. It was soon pointed out by Rasul and
Schlottmann\cite{Rasul} that these intriguing features can be
understood phenomenologically by an {\it spin-only} effective model
involving two impurity spins ${\vec S}_A$, ${\vec S}_B$, and two
conduction electron spins ${\vec s}_c(a)$, ${\vec s}_c(b)$, as shown
in Fig.\ref{picture}. As a result of many-body process, the
interaction terms in the corresponding Hamiltonian Eq.(\ref{h1})
emerge from the low energy regime of the original
TIKM\cite{JKW81,Affleck1,Affleck2}. The $T_K$ describes the
splitting between the Kondo singlet and the spin triplet states,
while the cross-coupling $K$ represents the interaction-induced
frustration. It is known that the critical point may be replaced by
a crossover in the absence of particle-hole
symmetry\cite{Affleck2,Sakai,Fye,Gan,Silva,Zarand}, while with this
symmetry a large degeneracy at the critical point is observed
\cite{Affleck2,Gan,note00}. Compatible with this observation, the
Rasul-Schlottmann (RS) spin model indeed exhibits an enhanced
degeneracy at a special point $P:(K/T_K=1,J_R/T_K=2)$ corresponding
to the critical point\cite{Rasul}.

On the other hand, Kondo effect or Kondo screening is conceptually
associated with the notion of "Kondo entanglement"
\cite{Hewson,Coleman1,Si1}. The Kondo models with a single or two
magnetic impurities have been investigated from the quantum
entanglement
perspective\cite{Costa,Hur1,Hur2,Affleck09,Costa2,Cho,Erik,Saleur}.
For the single impurity Kondo problem, Kondo screening implies
formation of a Kondo singlet groundstate\cite{Yosida}. This is an
entangled state consisting of the two parts, one is the local
magnetic impurity, another is the rest of the whole system. The
entanglement between the two parts defines the so-called
single-impurity Kondo entanglement (SIKE), a measure of Kondo
entanglement typically quantified by the von Neumann
entropy\cite{Bennetti}. The SIKE in a gapless bulk follows the
well-known scaling law of the thermodynamic entropy at large
distance over the coherent length of the Kondo
cloud\cite{Hur1,Affleck09}. For the TIKM, several different
impurity-related entanglements are considered. Similar to the SIKE,
one could consider the two-impurity Kondo entanglement (TIKE), i.e.,
the entanglement between the two impurities and the rest of the
system. The TIKE can be also measured by the von Neumann
entropy\cite{Bennetti,Cho}. Another useful quantity is the
inter-impurity entanglement (IIE), i.e., the entanglement between
the two local magnetic impurities\cite{Costa2,Cho,Bayat}. Such IIE
can be quantified by concurrence or negativity\cite{Wootters,Vidal}.
It has been shown that the IIE is non-zero when the RKKY interaction
is at least several times larger than the Kondo energy
scale\cite{Cho,Bayat}.
However, this feature alone does not sufficiently guarantee a true
phase transition\cite{Yang,note0} nor necessarily imply a full
suppression of the Kondo effect.

One should notice that the SIKE and TIKE quantified by the von
Neumann entropy in Kondo systems with two- or more impurities
usually mix the contributions from conduction electrons and other
impurities\cite{note1}, hence these quantities are not as distinct
as in the single impurity Kondo problem. Therefore, an alternative
measure of the Kondo entanglement, capable of characterizing the
variation of Kondo screening across the quantum critical point in
generic multi-impurity systems, is highly desirable\cite{note2}.

In this paper, we investigate a {\it local Kondo entanglement}
(LKE), namely, the entanglement between a magnetic impurity and its
near-by conduction electron only. The definition of the LKE in
generic Kondo systems is described in Appendix A. Generally,
suppression of this quantity should imply the complete destruction
of the Kondo effect, though its connection with the impurity quantum
phase transition as in the generic TIKM remains to be clarified. As
a concrete example,  the IIE and LKE in the RS model are evaluated
on equal-footing, both quantified by the concurrence or negativity.
An entanglement phase diagram is then obtained, exhibiting the
crossover from the Kondo singlet phase (with non-zero LKE) to the
inter-impurity AFM phase (with non-zero IIE). A critical point $P$
emerges along the strong frustration line $K=T_K$ where both LKE and
IIE show sudden changes and the groundstate wavefunction shows a
discontinuity.
In the following sections, we shall present the results of the exact
solutions of the RS model, the entanglement phase diagram, as well
as an entanglement monogamy. We also briefly discuss implications of
these results for generic Kondo lattices and quantum information
processing.

\section{Model and solutions}

\begin{figure}
\begin{center}
\includegraphics[width=0.30\textwidth]{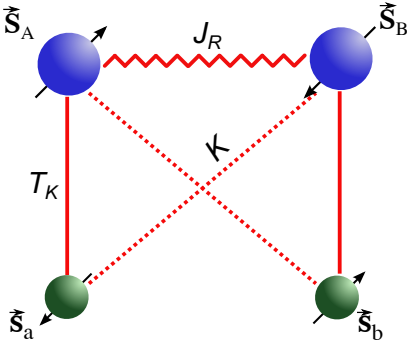}
\caption{(color online) The cartoon picture of the RS model: the
local magnetic moments and the conduction electron spins are denoted
by big blue circled and small green circled arrows, respectively.
$T_K$, $J_R$, and $K$ represent the single-ion Kondo energy scale,
the RKKY interaction, and the interaction induced cross Kondo
coupling. Various entanglements among these spins are defined in the
text.} \label{picture}
\end{center}
\end{figure}

The RS spin model, presumably taking into account the effective
many-body process, is a fixed-point Hamiltonian of the TIKM. It is
described by\cite{Rasul}
\begin{eqnarray}
H_{RS}&=&T_K[{\vec S}_A\cdot{\vec s}_c(a)+{\vec S}_B\cdot{\vec
s}_c(b)]+J_R
{\vec S}_A\cdot{\vec S}_B \nonumber\\
&& + ~K[{\vec S}_A\cdot{\vec s}_c(b)+{\vec S}_B\cdot{\vec s}_c(a)].
\label{h1}
\end{eqnarray} Where, ${\vec S}_A$ and ${\vec S}_B$ denote the spin
operators of the local moments at sites $A, B$, respectively. The
electron spin density at the spatial site $r=a,b$ is denoted by
${\vec
s}_c(r)=\frac{1}{2}\sum_{\alpha,\beta=\uparrow,\downarrow}c^{\dagger}_{\alpha}(r)
{\vec\sigma}_{\alpha\beta} c_{\beta}(r)$, with $c_{\alpha}(r)$ being
the annihilation operator of conduction electrons and ${\vec\sigma}$
the Pauli matrices. In the present two-impurity Kondo problem, only
the spin degrees of freedom are relevant while the charge degrees of
freedom are frozen. Thus ${\vec s}_c(a)$ and ${\vec s}_c(b)$ are
treated as two electron spins localized at the sites $r=a$ and $b$
respectively. The interaction parameters include $T_K$, the
single-ion Kondo temperature, $J_R$, the intersite RKKY interaction
energy, and $K$, the cross Kondo coupling between the local moments
and the electron spins, as shown in Fig.\ref{picture}\cite{Rasul}.
All these interaction terms are spin-$SU(2)$ invariant. The direct
Kondo coupling takes place between the impurity and conduction
electron in the assigned nearest neighboring sites $(A,a)$ or
$(B,b)$. The cross Kondo coupling term ($K$) is induced effectively
in the low energy limit of the original TIKM and $K\rightarrow T_K$
when the even/odd channel symmetry is imposed. For our purpose we
will assume that $J_{R}$ and $K$ are tunable independently with
respect to $T_K$. Because Eq. (1) is invariant under a combination
of the impurity permutation ${\vec S}_A\leftrightharpoons {\vec
S}_B$ and the exchange $ T_K\leftrightharpoons K$, the regime with
strong frustration corresponds to $K=T_K$. Hence we only need to
consider $0\leq K/T_K\leq 1$. In the following we set $T_K=1$
without losing generality.

The eigenstates of $H_{RS}$ can be classified into six catalogs: two
singlets, three triplets, and one quintet, according to the
decomposition of tensor representations of the $SU(2)$ group:
$\underline {\mathbf 2}\otimes \underline {\mathbf 2}\otimes
\underline {\mathbf 2} \otimes \underline {\mathbf 2}=2\times
\underline {\mathbf 1}\oplus 3\times\underline {\mathbf 3}\oplus
\underline {\mathbf 5}$ as listed in Appendix B. These eigenstates
are constructed based on a complete set of the conventional basis
$|S^z(A),s^z_c(a); S^z(B),s^z_c(b)\rangle$ and labeled by the total
spin $S$, its z-component $S_z$, and the parity (with respect to
permutations of the two local moments) \cite{Rasul}. Because we
consider $T_K$, $J_R$, and $K$ are all AFM, the groundstate is among
the mixed states of the two singlets of even parity. They are
denoted by (each up to a normalization factor)
\begin{eqnarray}
|\Psi_{(0^+)}\rangle&=&|\Downarrow\downarrow;\Uparrow\uparrow\rangle
+|\Uparrow\uparrow;\Downarrow\downarrow\rangle\\
&+&\frac{\Delta-J_R+2}{J_R-2K}(|\Downarrow\uparrow;\Downarrow\uparrow\rangle
+|\Uparrow\downarrow;\Uparrow\downarrow\rangle)\nonumber\\
&+&\frac{-\Delta+2K-2}{J_R-2K}(|\Downarrow\uparrow;\Uparrow\downarrow\rangle
+|\Uparrow\downarrow;\Downarrow\uparrow\rangle)\nonumber
\end{eqnarray}
and
\begin{eqnarray}
|\Psi_{(0'^+)}\rangle&=&|\Downarrow\downarrow;\Uparrow\uparrow\rangle
+|\Uparrow\uparrow;\Downarrow\downarrow\rangle\\
&+&\frac{-\Delta-J_R+2}{J_R-2K}(|\Downarrow\uparrow;\Downarrow\uparrow\rangle
+|\Uparrow\downarrow;\Uparrow\downarrow\rangle)\nonumber\\
&+&\frac{\Delta+2K-2}{J_R-2K}(|\Downarrow\uparrow;\Uparrow\downarrow\rangle
+|\Uparrow\downarrow;\Downarrow\uparrow\rangle)\nonumber.
\end{eqnarray}
The corresponding eigen energies are
\begin{eqnarray}
E_{(0^+)}&=&E_{(0)}-\frac{1}{2}\Delta,\\
E'_{(0^+)}&=&E_{(0)}+\frac{1}{2}\Delta,
\end{eqnarray}
with $E_{(0)}=-J_R/4-K/2-1/2$ and
$\Delta=\sqrt{(1+K-J_R)^2+3(K-1)^2}$.

Another relevant low energy states are from the odd triplet
\begin{eqnarray}
|\Psi_{(1^-,+1)}\rangle &=&
 \frac{J_R+\sqrt{(K-1)^{2}+J_R^{2}}}{1-K}
 (|\Downarrow\uparrow;\Uparrow\uparrow\rangle
-|\Uparrow\uparrow;\Downarrow\uparrow\rangle)\nonumber\\
&&-|\Uparrow\downarrow;\Uparrow\uparrow\rangle
+|\Uparrow\uparrow;\Uparrow\downarrow\rangle,\nonumber\\
|\Psi_{(1^-,~0)}\rangle &=&\frac{1-K+\sqrt{(K-1)^2+J_R^2}}{J_R}
(|\Downarrow\uparrow;\Uparrow\downarrow\rangle
-|\Uparrow\downarrow;\Downarrow\uparrow\rangle)\nonumber\\
 &&+|\Downarrow\downarrow;\Uparrow\uparrow\rangle
-|\Uparrow\uparrow;\Downarrow\downarrow\rangle,\nonumber\\
|\Psi_{(1^-,-1)}\rangle &=& \frac{J_R+\sqrt{(K-1)^{2}+J_R^{2}}}{1-K}
(|\Uparrow\downarrow;\Downarrow\downarrow\rangle
-|\Downarrow\downarrow;\Uparrow\downarrow\rangle)\nonumber\\
&&-|\Downarrow\uparrow;\Downarrow\downarrow\rangle
+|\Downarrow\downarrow;\Downarrow\uparrow\rangle.
\end{eqnarray}
The corresponding eigen energy is
\begin{eqnarray}
E_{(1^-)}=-\frac{J_R}{4}-\frac{\sqrt{(K-1)^2+J_R^2}}{2}.
\end{eqnarray}

It is apparent that the groundstate of the RS Hamiltonian
Eq.(\ref{h1}) is the singlet $|\Psi_{(0^+)}\rangle$. However, the
repulsive level spacing (energy gap) $\Delta$ vanishes at a special
point $P$, corresponding to $K=T_K=J_R/2$, so that the singlet
$|\Psi_{(0'^+)}\rangle$ is degenerate with $|\Psi_{(0^+)}\rangle$ at
this point. Meanwhile, the odd triplet $|\Psi_{(1^-)}\rangle$ is
degenerate with $|\Psi_{(0^+)}\rangle$ for $K=1, J_R\geq 2$. Hence
the model indeed shows a strong frustration along the line $K=1$ and
exhibits an enlarged symmetry at $P$\cite{note3}. The precise
wavefunctions across the $P$ point can be determined by taking
either limits ($K=1,J_R=2-\epsilon$) and ($K=1,J_R=2+\epsilon$),
$\epsilon\rightarrow 0^{+}$.  It readily reveals a discontinuity in
$|\Psi_{(0^+)}\rangle$ as shown in Appendix B.

\section {Entanglement phase diagram}

Now, we start from the conventional Kondo impurity entanglement,
i.e., the SIKE defined as the entanglement between a local moment,
say ${\vec S}(A)$, and the rest of the system, denoted by $\tilde
A$. It is measured by the van Norman entropy ${\cal
E}_{SIKE}=-Tr_{(A)}\{\hat{\rho}_{imp}(A)\ln{\hat\rho}_{imp}(A)\}$,
where ${\hat\rho}_{imp}(A)$ is the reduced density matrix
${\hat\rho}_{imp}(A)=Tr_{\tilde A} {\hat\rho}$. Here,
${\hat\rho}=|\Psi_G\rangle\langle\Psi_G|$ is the density matrix for
the groundstate ( in our case $|\Psi_G\rangle=|\Psi_{(0^+)}\rangle$)
of the whole system. It is straightforwardly seen that ${\cal
E}_{SIKE}=1$ due to the $SU(2)$-spin invariance, indicating a
maximal entanglement between a local moment and the reminder of the
whole system.

Next, we consider the TIKE, the entanglement of the two local
moments with the conduction electrons. Following Ref.\cite{Cho},
this entanglement is determined by the reduced density matrix of the
two impurities, ${\hat\rho}_{imp}(AB)=Tr_{(c)}{\hat\rho}$, with
$Tr_{(c)}$ indicating trace over the Hilbert subspace spanned by the
conduction electrons. It can be quantified by the von Neumann
entropy\cite{Cho}
\begin{eqnarray}
{\cal E}_{TIKE}=-p_s\log p_s-(1-p_s)\log \frac{1-p_s}{3}.
\end{eqnarray}
Here, $p_s=\frac{1}{4}-f_{AB}$ is the fidelity of the spin singlet
within the reduced two impurity state, and
$f_{AB}=\langle{\Psi_{0^+}}|{\vec S}(A)\cdot{\vec S}(B)
|{\Psi_{0^+}}\rangle$ is the spin-spin correlation function on the
groundstate. In our case,
\begin{eqnarray}
f_{AB}=\frac{1}{2}\frac{1+K-J_R}{\sqrt{(J_R-K-1)^2+3(K-1)^2}}-\frac{1}{4}.
\end{eqnarray}

\begin{figure}
\begin{center}
\includegraphics[width=0.45\textwidth]{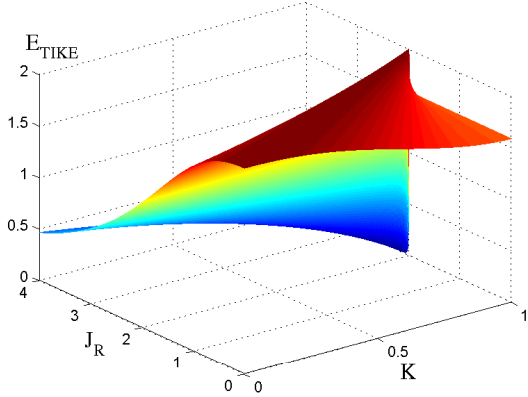}
\caption{(color online) The TIKE as a function of $J_R$ and $K$.}
\label{tike}\end{center}
\end{figure}

${\cal E}_{TIKE}$ is then evaluated on the groundstate
$|\Psi_{0^+}\rangle$  as shown in Fig.\ref{tike}. We find that
${\cal E}_{TIKE}$ is not only a smoothly varying function of
$f_{AB}$ as already shown in Ref.\cite{Cho}, but also a smooth
function of $K$ and $J_R$ without detectable feature across the
point $P$.

\begin{figure}
\begin{center}
\includegraphics[width=0.45\textwidth]{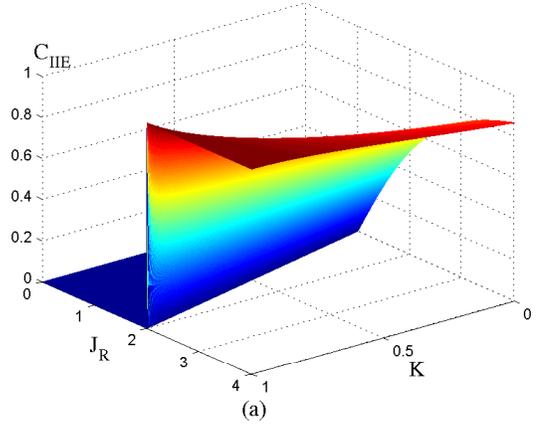}
\includegraphics[width=0.45\textwidth]{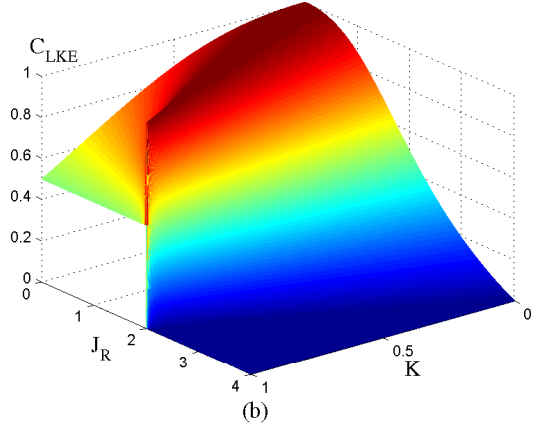}
\caption{(color online) The concurrence of  the IIE (a) and the LKE
(b) as functions of $J_R$ and $K$. The single-ion Kondo energy scale
is taken as unit $T_K=1$.} \label{concurrence}
\end{center}
\end{figure}

Now we turn to the IIE. It can be measured by the concurrence or
negativity , ${\cal C}_{IIE}$. Its evaluation is also related to the
reduced two-impurity density matrix ${\hat\rho}_{imp}(AB)$. The
concurrence can be expressed by\cite{Cho}
\begin{eqnarray}
{\cal C}_{IIE}=\max\{-2f_{AB}-1/2,0\}.
\end{eqnarray}
The result is plotted in Fig.\ref{concurrence}(a). For fixed $K<1$,
${\cal C}_{IIE}$ increases continuously with $J_R$. For $K=1$,
${\cal C}_{IIE}$ shows a sudden increase from zero to unity when
$J_R$ goes across $J_R=2$.

Together with the observed discontinuity in the groundstate
wavefunction, the jump in IIE evidences a transition at $P$
\cite{Wu}. But its relation with the suppression of Kondo effect
remains uncertain. We now turn to an alternative definition of the
Kondo entanglement, i.e., the LKE between a local moment, say ${\vec
S}(A)$, and the conduction electron at its nearest neighbor site,
${\vec s}_{c}(a)$. The LKE differs to the conventional impurity
entanglement as it involves only a spatially neighboring pair formed
by a local moment and a conduction electron. This local Kondo pair
is defined in the original real space point-contact Kondo
interaction $J_K {\vec S_{A}}\cdot {\vec s_c(a)}$ (with $J_K$ being
the original Kondo coupling) as shown in Appendix A. Similar to IIE,
the LKE can be evaluated by the concurrence or negativity, via the
corresponding reduced density matrix
${\hat\rho}_{LK}(Aa)=Tr_{({\tilde A}{\tilde a})}{\hat\rho}$, with
$Tr_{({\tilde A}{\tilde a})}$ indicating the trace in the Hilbert
space except the subspace spanned by ${\vec S}_A$ and ${\vec
s}_c(a)$. Thus we have
\begin{eqnarray}
{\cal C}_{LKE}=\max\{-2f_{Aa}-1/2,0\},
\end{eqnarray}
where $f_{Aa}=\langle{\Psi_{0^+}}|{\vec S}(A)\cdot{\vec s}_c(a)
|{\Psi_{0^+}}\rangle$ is the correlation function of the local Kondo
singlet state,
\begin{eqnarray}
f_{Aa}=\frac{1}{4}\frac{J_R+2K-4}{\sqrt{(J_R-K-1)^2+3(K-1)^2}}-\frac{1}{4}.
\end{eqnarray}
The result of ${\cal C}_{LKE}$ is plotted in
Fig.\ref{concurrence}(b). Interestingly, ${\cal C}_{LKE}$ develops a
maximum for $0<K<1$, $J_R<2$ and decreases monotonically for
$J_R>2$. But along the line $K=1$ it shows a sudden suppression
$J_R>2$ .

\begin{figure}
\begin{center}
\includegraphics[width=0.35\textwidth]{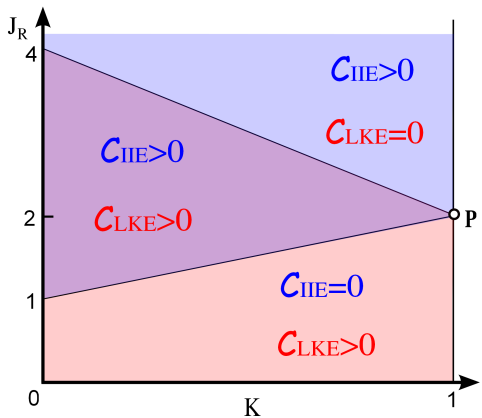}
\caption{(color online) Entanglement phase diagram: there are three
distinct phases divided by the lines ${\cal C}_{LKE}=0$ and ${\cal
C}_{IIE}=0$. These two lines intersect at the singular point
$P:(K=1, J_R=2)$, where the coexistence regime diminishes leading to
a Kondo breakdown transition to the AFM singlet phase.}
\label{diagram}
\end{center}
\end{figure}

An entanglement phase diagram in terms of $K$ and $J_R$ is then
drawn in Fig.\ref{diagram}, where three different phases divided by
the lines $J_R=K+1$ and $J_R=4-2K$ are indicated: the IIE phase
(${\cal C}_{IIE}>0$,${\cal C}_{LKE}=0$), the LKE phase (${\cal
C}_{LKE}>0$,${\cal C}_{IIE}=0$), and the co-existence  phase (${\cal
C}_{IIE}>0$, ${\cal C}_{LKE}>0$). The IIE and LKE phases contact
only at the point $P$: by increasing $J_R$ across $P$ along the
strong frustration line $K=1$, $f_{AB}$ has a sudden drop from $1/4$
to $-1/2$, $f_{Aa}$ has a jump from $-1/2$ to $0$. Or, ${\cal
C}_{IIE}$ has a jump from $0$ to $1$ while ${\cal C}_{LKE}$ has a
sudden drop from $1/2$ to $0$.

Therefore, together with the discontinuity of the wavefunction
$|\Psi_{(0+)}\rangle $, the sudden changes along the line $K=1$ in
the IIE and LKE do evidence a phase transition accompanied by a
breakdown of Kondo effect. Of course, a true second order phase
transition usually involves a continuous variation of the order
parameter before its suppression. So the discontinuity exhibited in
the IIE or LKE ( as an order parameter here) is seemingly due to the
simplicity of the present model involving only two conduction
electrons.

We notice that the inter-impurity spin-spin correlation function,
closely related to the IIE discussed here, was actually calculated
in the early numerical renormalization group study \cite{Jones88} at
low yet finite temperatures, where the calculated quantity did not
exhibit a sudden discontinuity but a rather sharp change at the
critical point. Interestingly, a more recent calculation based on
the natural orbitals renormlization group method does find a suddent
jump of this quantity at zero temperture\cite{He15}. Therefore, the
discontinuity of the IIE and LKE may be not limited to the present
model, but a generic feature of the impurity state involving finite
degrees of freedom at quantum critical points. This discontinuity
could be smeared when the impurity degrees of freedom become
infinite as in the Kondo lattice case.

\section{The fidelity of the Kondo and inter-impurity
AFM singlets}

In order to clarify whether the phase with non-zero ${\cal C}_{IIE}$
or ${\cal C}_{LKE}>0$ corresponds to the AFM inter-impurity or Kondo
singlets, we calculate the normalized wavefunction overlaps $\langle
\Psi_{AFM}| \Psi_G\rangle $ and $\langle \Psi_{KS}|\Psi_G\rangle$,
respectively. We consider the pure state of the Kondo screening
phase (denoted by $|\Psi_{KS}\rangle$) as the groundstate at the
fixed point $K=J_R=0$. Similarly, we denote $ |\Psi_{AFM}\rangle $
the pure inter-impurity AFM state at the fixed point $K=0,
J_R\rightarrow \infty$. Here,
$|\Psi_{KS}\rangle=|\Psi_{(0)Aa}\rangle\otimes|\Psi_{(0)Bb}\rangle$,
$|\Psi_{AFM}\rangle=|\Psi_{(0)AB}\rangle\otimes|\Psi_{(0)ab}\rangle$,
with $|\Psi_{(0)Aa}\rangle=\frac{1}{\sqrt 2}
(|\Uparrow\downarrow\rangle-|\Downarrow\uparrow\rangle)$,
$|\Psi_{(0)Bb}\rangle=\frac{1}{\sqrt 2}
(|\Uparrow\downarrow\rangle-|\Downarrow\uparrow\rangle)$,
$|\Psi_{(0)AB}\rangle=\frac{1}{\sqrt 2}
(|\Uparrow\Downarrow\rangle-|\Downarrow\Uparrow\rangle)$,
$|\Psi_{(0)ab}\rangle=\frac{1}{\sqrt 2}
(|\uparrow\downarrow\rangle-|\downarrow\uparrow\rangle)$.

We find that in the regime with ${\cal C}_{IIE}=0$ or ${\cal
C}_{LKE}=0$ the respective wavefunction overlap vanishes. Therefore,
the obtained entanglement phase diagram Fig.\ref{diagram} reflects
the overall evolution from the Kondo singlet to the inter-impurity
AFM singlet.

\begin{figure}
\begin{center}
\includegraphics[width=0.45\textwidth]{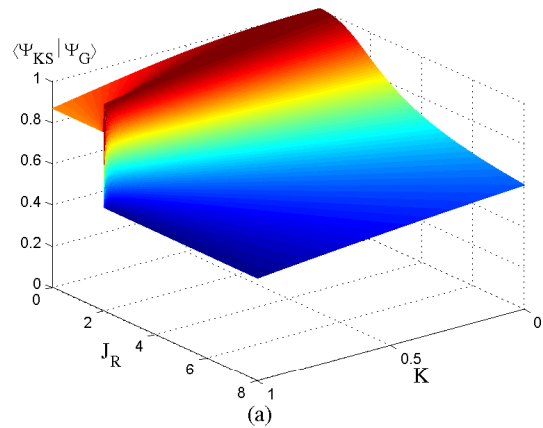}
\includegraphics[width=0.45\textwidth]{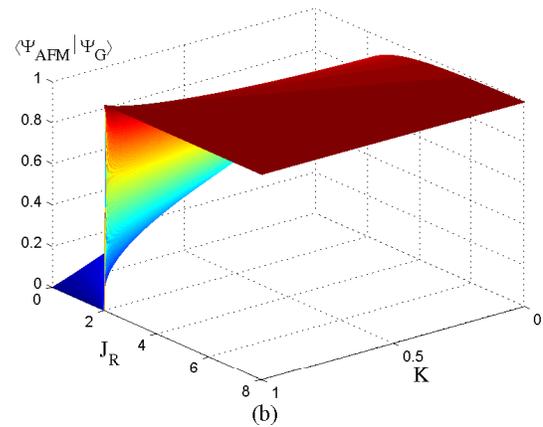}
\caption{(color online) (a) The fidelity of the Kondo singlet
$|\Psi_{KS}\rangle$ on the groundstate $|\Psi_{G}\rangle $. (b) The
fidelity of the inter-impurity AFM singlet $ |\Psi_{AFM}\rangle $ on
the groundstate $|\Psi_{G}\rangle $.}\label{fidelity}
\end{center}
\end{figure}

\begin{figure}
\begin{center}
\includegraphics[width=0.45\textwidth]{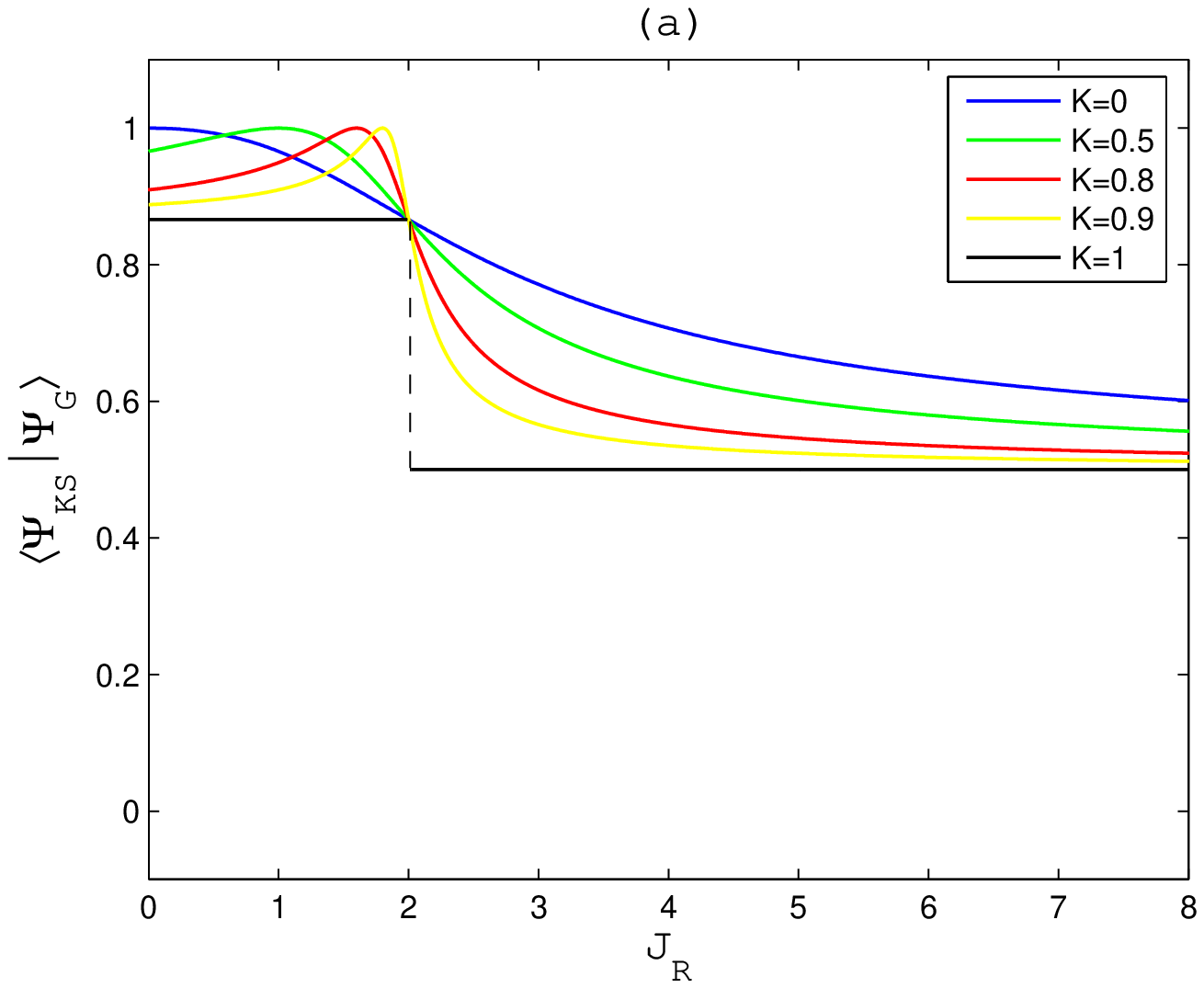}
\includegraphics[width=0.45\textwidth]{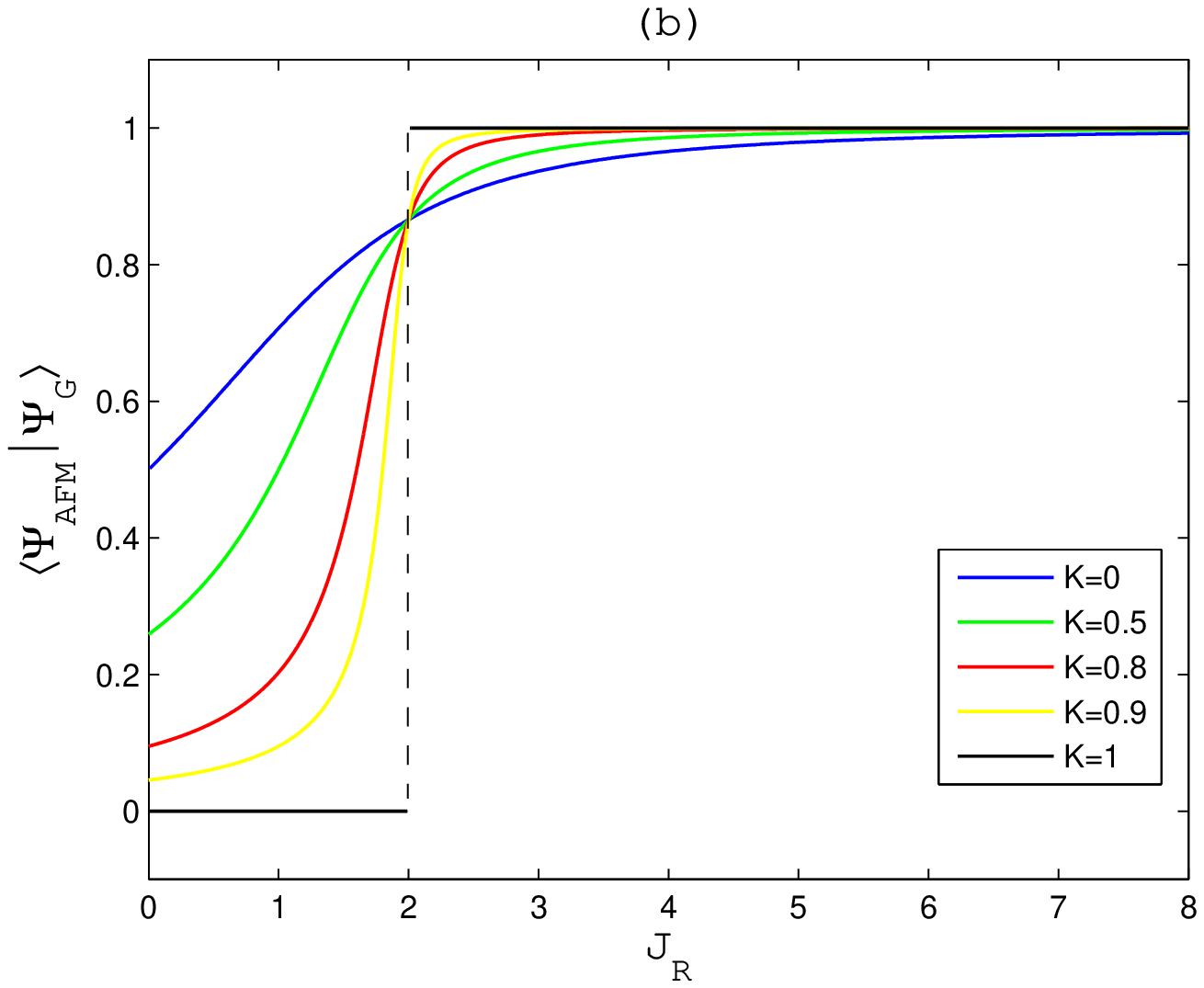}
\caption{(color online) The groundstate fidelity of the Kondo
singlet $|\Psi_{KS}\rangle$ (a) or of the inter-impurity AFM singlet
(b) as a function of the RKKY interaction for various fixed
frustration $K$.}\label{fidelity2}
\end{center}
\end{figure}

Specifically, increasing $K$ will enhance the frustration, so that
the fidelities of $|\Psi_{KS}\rangle$ and $ |\Psi_{AFM}\rangle $ on
the true groundstate $ |\Psi_{G}\rangle $ change with varying $J_R$.
The fidelities  can be calculated as the normalized wavefunction
overlaps $\langle \Psi_G|\Psi_{AFM} \rangle $ and $\langle
\Psi_G|\Psi_{KS}\rangle$. The three-dimensional plots of the
wavefunctions overlaps are shown in Fig.\ref{fidelity}(a) and Fig.
\ref{fidelity}(b), respectively. Fig.\ref{fidelity2}(a) and
Fig.\ref{fidelity2}(b) show the $J_R$-dependence of these quantities
for fixed values of $K$. We find that $\langle
\Psi_G|\Psi_{AFM}\rangle$ increases rapidly with $J_R$ and saturates
to 1 after entering the AFM phase, $J_R>2$. Notice that $\langle
\Psi_G|\Psi_{KS}\rangle$ always approaches to $1/2$ for
$J_R\rightarrow \infty$, indicating that the Kondo singlet fidelity
$p_s\rightarrow 1/2$. As shown in Appendix A, this implies the
non-correlated Kondo state or a full suppression of Kondo effect.

\section{Entanglement sum rule}

By definition, the concurrences of LKE and IIE introduced in the
main text, ${\cal C}_{IIE}$ and ${\cal C}_{LKE}$, are always
non-zero ( but $\leq 1$) when the respective correlation function
$f_{AB}\leq -1/4$ or $f_{Aa}\leq -1/4$. Based on the exact
solutions, these functions can be reexpressed as
\begin{eqnarray}
f_{AB}=-\frac{\lambda}{2}\cos\theta-\frac{1}{4};
f_{Aa}=\frac{\lambda}{4}\cos\theta-\frac{\sqrt 3}{4}\sin\theta
-\frac{1}{4}.
\end{eqnarray}
Where, $\theta=\tan^{-1}\frac{{\sqrt 3}(1-K)}{|J_R-K-1|}$,
$\lambda=\frac{J_R-K-1}{|J_R-K-1|}$. Therefore, in the whole
parameter regime $J_R\geq 0$, $K\leq 1$, or $0\leq \theta\leq
\frac{\pi}{2}$, we have
\begin{eqnarray}
{\cal C}_{IIE}&=&\max\{\lambda\cos\theta,0\};\\
{\cal C}_{LKE}&=&\max\{-\frac{\lambda}{2}\cos\theta+\frac{\sqrt
3}{2}\sin\theta,0\}.
\end{eqnarray}
Thus ${\cal C}_{IIE}$ and ${\cal C}_{LKE}$ are both non-zero when
$K+1<J_R<4-2K$. In this co-existence regime where $\frac{\pi}{6}<
\theta<\frac{\pi}{2}$, we have
\begin{eqnarray}
{\cal C}_{IIE}+ {\cal C}_{LKE}=\sin(\theta+\frac{\pi}{6}).
\end{eqnarray}
This is a sum rule constraining the variations of ${\cal C}_{IIE}$
and ${\cal C}_{LKE}$: $\frac{\sqrt 3}{2}<{\cal C}_{IIE}+ {\cal
C}_{LKE}<1$ in this regime.

As an inference of the sum rule, we note that a Werner state with
$p_s\geqq (1+3/\sqrt 2)/4\approx 0.78$ has the nonlocal correlation
characteristic, i.e., violation of the Bell
inequality\cite{Cho,Clauser,Horodecki}. This value corresponds to $
{\cal C}\geqq 0.56$. Therefore, we have an entanglement monogamy:
${\cal C}_{LKE}$ and ${\cal C}_{IIE}$ cannot simultaneously
maximize, nor even simultaneously fall into this region. In other
words, only one of the entanglements could be maximized or violate
the Bell inequality.

Finally, one can also define the entanglement between the spin
${\vec S}_A$ and ${\vec s}_c (b)$, i.e., the cross Kondo
entanglement, quantified by ${\cal C}_{CKE}$. Owing to the discrete
symmetry mentioned previously, this quantity can be drived similar
to ${\cal C}_{LKE}$, obtaining ${\cal
C}_{CKE}=\max\{-\frac{\lambda}{2}\cos\theta-\frac{\sqrt
3}{2}\sin\theta,0\}$. So ${\cal C}_{CKE}$ is non-zero only when
$J_R<4K-2$. This region is a subregion of the LKE phase, below the
straight line connecting ($K=0.5, J_R=0$) and the $P$-point(not
shown in Fig.\ref{diagram} ). Hence including ${\cal C}_{CKE}$ does
not violate the previous sum rule. Moreover, another sum rule ${\cal
C}_{LKE}+ {\cal C}_{CKE}=\cos\theta$ holds in this subregion.
Therefore the previous inference as well as all the conclusions in
the main text remain valid in the whole phase diagram.

\section{Implications and discussions}

As we emphasized, the quantum entanglement related to the single-ion
Kondo effect is complicated when two or more local magnetic
impurities are introduced. We have considered the LKE in the case of
two magnetic impurities where a quantum phase transition from the
Kondo to the AFM singlets takes place. The exact results obtained
from a simplified yet effective TIKM, i.e., the RS spin model, show
a phase diagram involving several regimes corresponding to non zero
LKE and IIE, and their co-existence. With increasing frustrated
cross coupling the co-existence regime shrinks and vanishes at the
critical point $P$ where the groundstate $|\Psi_{G}\rangle$ has a
discontinuity accompanied by jumps in the IIE and LKE. The
discontinuity of such quantities at the critical point may be a
generic feature caused by the non-extensive impurity term in the
free energy of the TIKM\cite{Bayat14,He15}.

Due to the entanglement sum rule, the LKE and IIE cannot be
simultaneously maximized even in the co-existence regime. Because an
entangled state with non-zero concurrence while still keeping the
Bell-CHSH inequality\cite{Clauser,Horodecki} can be used for quantum
information processing including quantum
teleportation\cite{Popescu,Lee}, the singlet groundstate in the
co-existence phase shares this feature either by the LKE or IIE, and
thus should be an interesting candidate state for quantum
information processing.

It is interesting to understand implications of our present study
for more generic Kondo lattice models. In these systems, like the
heavy fermion metals, the magnetic quantum phase transitions may be
influenced by the variation of Kondo effect. A local quantum phase
transition is possible in more generic Kondo lattice phase diagram
where the criticality is associated with a critical breakdown of the
single-ion Kondo effect\cite{Si2,Coleman2,Si3}. Near the critical
point, the Hall constant shows a discontinuous jump due to the
reconstruction of the Fermi surface across the critical
point\cite{Si2,Coleman2}. Experimental evidence for this scenario
comes from several prototypes of heavy fermion metals including
YbRh$_2$Si$_2$\cite{Paschen04} and CeNiAsO\cite{Luo} where the
observed Hall constant exhibits a sudden change accompanying the
magnetic phase transition.

Although the above scenario could be naturally understood based on
the Kondo entanglement picture, a lattice model Hamiltonian with
exact solutions showing the Kondo entanglement breakdown transition
is lacking. Thus our studied model can serve as a toy model to
understand its basic physics from the quantum entanglement
perspective. On the one hand, one expects that with the evolution
from two impurities to an regular local moment lattice, the Kondo
singlet and inter-impurity AFM singlet states evolve into the
paramagnetic heavy fermion and AFM ordered phases, respectively. On
the other hand, in addition to the RKKY interaction, the crossing
Kondo coupling ($K$) emerges as a many-body frustration effect and
plays a role in controlling the transition. Generally, in the regime
with relatively small $K$, the Kondo and inter-impurity singlets can
co-exist in the intermediate regime of $J_R$. The co-existence
regime diminishes with increasing $K$ and a direct Kondo singlet
breakdown transition takes place when the Kondo coupling is
maximally frustrated (at $K=T_K$). In the Kondo lattice phase
diagram this condition should correspond to the regime with strong
geometric frustrations and quantum fluctuations but no spin liquid
phase sets in\cite{Si3}. Finally, extending the present study to the
Kondo lattice cases with frustrated cross couplings or various
anisotropic interactions is not an easy task but highly desirable.
Nevertheless, our present result already provides a concrete example
of Kondo breakdown quantum phase transition from the quantum
entanglement perspective.

\section*{Acknowledgments}

One of the authors (J.D.) would like to thank Rong-Qiang He,
Zhong-Yi Lu, and Qimiao Si for useful discussions. He especially
thanks Rong-Qiang He and Zhong-Yi Lu for identifying the nature of
discontinuity of the inter-impurity spin-spin correlation function.
This work was supported in part by the NSF of China under Grant No.
11304071 and No. 11474082.

\appendix

\section{The LKE in the $N$-impurity Kondo system}

The Kondo system consisting of a metallic electron host, with energy
dispersion $\epsilon({\vec k})$, and $N$-number of quantum
mechanical magnetic moments (of spin-1/2) localized at sites $i$,
$i=1,2, ......,N$, is described by the Hamiltonian
\begin{eqnarray}\label{nkondo}
H_{NKM}&=&\sum_{\sigma=\uparrow,\downarrow}\int d^3 {\vec
k}~\epsilon({\vec k})c^{\dagger}_{\sigma}({\vec k})c_{\sigma}({\vec
k})\nonumber\\
&&+J_K\sum^{N}_{i=1}{\vec S}_{i}\cdot{\vec s}_c({\vec r}_i),
\end{eqnarray}
where the Kondo coupling is either {\it antiferromagnetic} ($J_K>0$)
and {\it local}, in the sense that it takes place between a local
moment ${\vec S}_{i}$ and a spatially nearby conduction electron at
site ${\vec r}_i$ via the point-like interaction  $J_K {\vec
S}_{i}\cdot{\vec s}_c({\vec r}_i)$, with the electron spin operator
${\vec s}_c({\vec r}_i)=\frac{1}{2}c^{\dagger}_{\sigma}({\vec r}_i)
{\vec \sigma}_{\sigma\sigma'}c_{\sigma'}({\vec r}_i)$.

The local Kondo entanglement (LKE) refers to the quantum
entanglement of a local Kondo pair consisting of an impurity fixed
spin ${\vec S}_i$ and its spatially nearby conduction electron. As
we focus on the spin degrees of freedom relevant in the Kondo
screening, such a pair of spins is dubbed as a local Kondo state. It
constitutes a subsystem, with a local Hilbert space $\Omega_{i}$
spanned by ${\vec S}_i$ and ${\vec s}_c({\vec r}_i)$. Let $\Omega$
be the total Hilbert space, and ${\tilde \Omega_{i}}$ the Hilbert
subspace complementary to $\Omega_{i}$: $\Omega_{i}\cup
{\tilde\Omega_{i}}=\Omega$. Assume $|\Psi_G\rangle$ be the
groundstate of the $N$-impurity Kondo model $H_{NKM}$ defined in
Eq.(\ref{nkondo}), ${\hat\rho}=|\Psi_G\rangle\langle\Psi_G|$ the
corresponding density matrix of the whole system, the LKE is defined
as the entanglement of the reduced density matrix of this subsystem
obtained by taking trace over all other degrees of freedom except
the local Hilbert space: $ {\hat\rho}_{\Omega_i}=Tr_{{\tilde
\Omega}_i}{\hat \rho} $. Obviously, $\Omega_i$ has dimension 4, so
${\hat\rho}_{\Omega_i}$ can be expressed by the matrices $I_{4\times
4}$, ${\vec\tau}\otimes{\vec\sigma}$, with ${\vec\tau}$ being the
Pauli matrices of the local spin, ${\vec
S}_i=\frac{1}{2}{\vec\tau}_{i}$\cite{Nielson}. Owing to the fact
that the Hamiltonian is real, $SU(2)$ invariant, and the groundstate
is always a spin-singlet, one has the general
form\cite{Nielson,Amico}
\begin{eqnarray}
{\hat\rho}_{\Omega_i}=\frac{1}{4}I_{4\times
4}+\frac{r}{4}[{\tau^{x}_i\otimes\sigma^{x}_i}
+{\tau^{y}_i\otimes\sigma^{y}_i} +{\tau^{z}_i\otimes\sigma^{z}_i}].
\end{eqnarray}

In terms of the Bell basis, the maximal entangled states
$|\Psi^{(\pm)}\rangle=\frac{1}{\sqrt 2} (|\Uparrow\downarrow\rangle
\pm |\Downarrow\uparrow\rangle )$ and
$|\Phi^{(\pm)}\rangle=\frac{1}{\sqrt 2} (|\Uparrow\uparrow\rangle
\pm |\Downarrow\downarrow\rangle )$, the reduced density matrix can
be expressed by
\begin{eqnarray}
{\hat\rho}_{\Omega_i}&=&p_s |\Psi^{(-)}\rangle\langle\Psi^{(-)}|\\
&+&p_t( |\Psi^{(+)}\rangle\langle\Psi^{(+)}|+
|\Phi^{(+)}\rangle\langle\Phi^{(+)}|+|\Phi^{(-)}\rangle\langle\Phi^{(-)}|),\nonumber
\end{eqnarray}
with $p_s$ and $p_t$ being the probabilities of spin singlet and
spin-triplet, respectively. Thus, ${\hat\rho}_{\Omega_i}$ is a
mixture of spin-singlet and spin-triplet. $p_s$ and $p_t$ are
related to the spin-spin correlation function $f_s=\langle
\Psi_G|{\vec S}_i\cdot{\vec s}({\vec r}_i)|\Psi_G\rangle$ via
$p_s=\frac{1}{4}-f_s$,$p_t=\frac{1}{4}+\frac{f_s}{3}$. Therefore,
the pure local Kondo singlet corresponds to $p_s=1$ or
$f_s=-\frac{3}{4}$, while the pure local Kondo triplet corresponds
to $p_t=\frac{1}{3}$ or $f_s=\frac{1}{4}$. Notice that the case of
$f_s=-\frac{1}{4}$ or $p_s=p_t=1/2$ corresponds to the
non-correlated Kondo pair with equal mixture of spin-singlet and
spin triplet.

In our definition, the LKE is measured by the concurrence of
${\hat\rho}_{\Omega_i}$. Because ${\hat\rho}_{\Omega_i}$ can be
reexpressed as $ {\hat\rho}_{\Omega_i}=\frac{4p_s-1}{3}
|\Psi^{(-)}\rangle\langle\Psi^{(-)}|+\frac{1-p_t}{3}I_{4\times 4}$,
it is also a Werner state\cite{Werner} with the spin singlet
fidelity $p_s=\langle
\Psi^{(-)}|{\hat\rho}_{\Omega_i}|\Psi^{(-)}\rangle$. The the
concurrence of this state is given by\cite{Wootters}
\begin{eqnarray}
{\cal C}_{LKE}=\max\{2p_s-1,0\}.
\end{eqnarray}
Such entanglement can be also measured by the negativity\cite{Vidal}
of the reduced density matrix, ${\cal N}_{{\hat\rho}_{\Omega_i}}$.
It is straightforwardly seen that the negativity of the Werner state
is equal to the concurrence\cite{Lee}, i.e., ${\cal
N}_{{\hat\rho}_{\Omega_i}}={\cal C}_{LKE}$.

Similarly, the entanglement of a pair of two local magnetic moments,
say ${\vec S}_i$ and ${\vec S}_j$, can be defined following the
previous approach. Parallel to the LKE, such the inter-impurity
entanglement (IIE) measured by the concurrence (${\cal C}_{IIE}$) is
closely related to the spin-spin correlation function $\langle
\Psi_G|{\vec S}_i\cdot{\vec S}_j|\Psi_G\rangle$. This quantity
varies monotonically with the RKKY interaction $J_R(ij)$. The {\it
indirect} RKKY interaction, i.e., the coupling between the local
moments which is not explicitly present in the bare Hamiltonian, is
generated in the second order perturbtion in $J_K$ and dependend on
the density of states at the Fermi energy and the spatial separation
$|{\vec R}_i-{\vec R}_j|$ of the two local moments. Because
$J_R(ij)$ oscillates non-universally, the IIE's of different pairs
$(i,j)$ are usually complicated. Remarkably, there are two special
situations where the IIE competes with the LKE: (i) the two impurity
case with $N=2$; (ii) The sites $i$ and $j$ are the
nearest-neighboring sites in the Kondo lattice with $N=L$ ($L=$ the
total number of sites). Therefore, the competition between the LKE
and IIE manifests the competition between the Kondo singlet and the
inter-impurity AFM singlet in the TIKM, or manifests the competition
between the paramagnetic heavy Fermion state and the AFM ordered
state in the Kondo lattices.

A phenomenological understanding of above competition invokes two
energy scales, the single-ion Kondo temperature $T_K\sim
D\sqrt{\rho_F J_K}\exp [-\frac{1}{\rho_F J_K}]$, and the RKKY
interaction $J_R$, with $D$ the band width and $\rho_F$ the density
of states at the Fermi energy. In the renormalization group
treatment one approaches the low energy limit by gradually
integrating out the degrees of freedom of conduction electrons. So
that at $T=0$, the latter can be introduced by adding a
 {\it direct} RKKY term $J_R {\vec S}_i\cdot {\vec
S}_j$ into the original Hamiltonian, while the single-ion Kondo
screening effect is described by a  term $T_K {\vec S}_i\cdot {\vec
s}_c({\vec r}_i)$. Phenomenologically, $T_K$ is the excitation
energy of the Kondo triplet above the Kondo singlet groundstate. The
effective cross Kondo coupling (denoted by $K$ in the main text)
between ${\vec S}_j$ and ${\vec s}_c({\vec r}_i)$ can be also
induced by pure quantum mechanical many-body processes. Proper
inter-impurity distance and the direct RKKY interaction guarantee
the required particle-hole symmetry\cite{Affleck2}. These
interpretations provide a basis for the RS spin model as a minimal
fixed point Hamiltonian of the TIKM.

\section{Eigenstates and eigen energies of the Rasul-Schlottmann
model}

The sixteen eigenstates (each up to a normalization factor) and the
corresponding eigenvalues are solved as following:
\begin{eqnarray}
|\Psi_{(0^+)}\rangle
&=&|\Downarrow\downarrow;\Uparrow\uparrow\rangle
+|\Uparrow\uparrow;\Downarrow\downarrow\rangle
+c_1(|\Downarrow\uparrow;\Downarrow\uparrow\rangle
+|\Uparrow\downarrow;\Uparrow\downarrow\rangle)\nonumber\\
&&-(c_1+1)(|\Downarrow\uparrow;\Uparrow\downarrow\rangle
+|\Uparrow\downarrow;\Downarrow\uparrow\rangle)
\end{eqnarray}

\begin{eqnarray}
E_{(0^+)}&=&-\frac{J_R}{4}-\frac{K}{2}-\frac{1}{2}\nonumber\\
&&-\frac{\sqrt{(1+K-J_R)^2+3(K-1)^2}}{2}
\end{eqnarray}

\begin{eqnarray}
|\Psi_{(0'^+)}\rangle
&=&|\Downarrow\downarrow;\Uparrow\uparrow\rangle
+|\Uparrow\uparrow;\Downarrow\downarrow\rangle +
c_2(|\Downarrow\uparrow;\Downarrow\uparrow\rangle
+|\Uparrow\downarrow;\Uparrow\downarrow\rangle)\nonumber\\
&&-(c_2+1)(|\Downarrow\uparrow;\Uparrow\downarrow\rangle
+|\Uparrow\downarrow;\Downarrow\uparrow\rangle)
\end{eqnarray}

\begin{eqnarray}
E_{(0'^+)}&=&-\frac{J_R}{4}-\frac{K}{2}\nonumber\\
&&-\frac{1}{2}+\frac{\sqrt{(1+K-J_R)^2+3(K-1)^2}}{2}
\end{eqnarray}

\begin{eqnarray}
|\Psi_{(1^+,+1)}\rangle&=&
 |\Downarrow\uparrow;\Uparrow\uparrow\rangle
-|\Uparrow\downarrow;\Uparrow\uparrow\rangle
+|\Uparrow\uparrow;\Downarrow\uparrow\rangle
-|\Uparrow\uparrow;\Uparrow\downarrow\rangle\nonumber\\
|\Psi_{(1^+,~0)}\rangle&=&
 |\Downarrow\uparrow;\Downarrow\uparrow\rangle
-|\Uparrow\downarrow;\Uparrow\downarrow\rangle\\
|\Psi_{(1^+,-1)}\rangle&=&
 |\Uparrow\downarrow;\Downarrow\downarrow\rangle
-|\Downarrow\uparrow;\Downarrow\downarrow\rangle
+|\Downarrow\downarrow;\Uparrow\downarrow\rangle
-|\Downarrow\downarrow;\Downarrow\uparrow\rangle\nonumber
\end{eqnarray}

\begin{eqnarray}
E_{(1^+)}=\frac{J_R}{4}-\frac{K}{2}-\frac{1}{2}
\end{eqnarray}

\begin{eqnarray}
|\Psi_{(1^-,+1)}\rangle &=&
 c_3(|\Downarrow\uparrow;\Uparrow\uparrow\rangle
-|\Uparrow\uparrow;\Downarrow\uparrow\rangle)\nonumber\\
&&-|\Uparrow\downarrow;\Uparrow\uparrow\rangle
+|\Uparrow\uparrow;\Uparrow\downarrow\rangle\nonumber\\
|\Psi_{(1^-,~0)}\rangle &=&
 |\Downarrow\downarrow;\Uparrow\uparrow\rangle
-|\Uparrow\uparrow;\Downarrow\downarrow\rangle\nonumber\\
 &&+c_4
(|\Downarrow\uparrow;\Uparrow\downarrow\rangle
-|\Uparrow\downarrow;\Downarrow\uparrow\rangle)\\
|\Psi_{(1^-,-1)}\rangle &=&
c_3(|\Uparrow\downarrow;\Downarrow\downarrow\rangle
-|\Downarrow\downarrow;\Uparrow\downarrow\rangle)\nonumber\\
&&-|\Downarrow\uparrow;\Downarrow\downarrow\rangle
+|\Downarrow\downarrow;\Downarrow\uparrow\rangle\nonumber
\end{eqnarray}

\begin{eqnarray}
E_{(1^-)}=-\frac{J_R}{4}-\frac{\sqrt{(K-1)^2+J_R^2}}{2}
\end{eqnarray}

\begin{eqnarray}
|\Psi_{(1'^-,+1)}\rangle &=&
 c_5(|\Downarrow\uparrow;\Uparrow\uparrow\rangle
-|\Uparrow\uparrow;\Downarrow\uparrow\rangle)\nonumber\\
&&-|\Uparrow\downarrow;\Uparrow\uparrow\rangle
+|\Uparrow\uparrow;\Uparrow\downarrow\rangle\nonumber\\
|\Psi_{(1'^-,~0)}\rangle &=&
 |\Downarrow\downarrow;\Uparrow\uparrow\rangle
-|\Uparrow\uparrow;\Downarrow\downarrow\rangle\nonumber\\
&&+c_6(|\Downarrow\uparrow;\Uparrow\downarrow\rangle
-|\Uparrow\downarrow;\Downarrow\uparrow\rangle)\\
|\Psi_{(1'^-,-1)}\rangle &=&
c_5(|\Uparrow\downarrow;\Downarrow\downarrow\rangle
-|\Downarrow\downarrow;\Uparrow\downarrow\rangle)\nonumber\\
&&-|\Downarrow\uparrow;\Downarrow\downarrow\rangle
+|\Downarrow\downarrow;\Downarrow\uparrow\rangle\nonumber
\end{eqnarray}

\begin{eqnarray}
E_{(1'^-)}=-\frac{J_R}{4}+\frac{\sqrt{(K-1)^2+J_R^2}}{2}
\end{eqnarray}

\begin{eqnarray}
|\Psi_{(2,~2)}\rangle &=&
 |\Uparrow\uparrow;\Uparrow\uparrow\rangle\nonumber\\
|\Psi_{(2,~1)}\rangle &=&
|\Downarrow\uparrow;\Uparrow\uparrow\rangle
+|\Uparrow\downarrow;\Uparrow\uparrow\rangle
+|\Uparrow\uparrow;\Downarrow\uparrow\rangle
+|\Uparrow\uparrow;\Uparrow\downarrow\rangle\nonumber\\
|\Psi_{(2,~0)}\rangle &=&
|\Downarrow\downarrow;\Uparrow\uparrow\rangle
+|\Downarrow\uparrow;\Downarrow\uparrow\rangle
+|\Downarrow\uparrow;\Uparrow\downarrow\rangle
+|\Uparrow\downarrow;\Downarrow\uparrow\rangle\nonumber\\
&&+|\Uparrow\downarrow;\Uparrow\downarrow\rangle
+|\Uparrow\uparrow;\Downarrow\downarrow\rangle\\
 |\Psi_{(2,-1)}\rangle &=&
|\Uparrow\downarrow;\Downarrow\downarrow\rangle
+|\Downarrow\uparrow;\Downarrow\downarrow\rangle
+|\Downarrow\downarrow;\Uparrow\downarrow\rangle
+|\Downarrow\downarrow;\Downarrow\uparrow\rangle\nonumber
\\
|\Psi_{(2,-2)}\rangle &=&
 |\Downarrow\downarrow;\Downarrow\downarrow\rangle\nonumber
\end{eqnarray}

\begin{eqnarray}
E_{(2^+)}=\frac{J_R}{4}+\frac{K}{2}+\frac{1}{2}
\end{eqnarray}

In above,
$c_1=\frac{\sqrt{(1+K-J_R)^{2}+3(K-1)^{2}}-J_R+2}{J_R-2K}$,
$c_2=\frac{-\sqrt{(1+K-J_R)^{2}+3(K-1)^{2}}-J_R+2}{J_R-2K}$,
$c_3=\frac{J_R+\sqrt{(K-1)^{2}+J_R^{2}}}{1-K}$,
$c_4=\frac{1-K+\sqrt{(K-1)^2+J_R^2}}{J_R}$,
$c_5=\frac{J_R-\sqrt{(K-1)^{2}+J_R^{2}}}{1-K}$, and
$c_6=\frac{1-K-\sqrt{(K-1)^2+J_R^2}}{J_R}$. Along the line $K=1$ or
$J_R=2K$, some coefficients are divergent, but the correct forms can
be obtained by taking the limits from either sides. In particular,
in the vicinity of the singular point $P:(K=1, J_R=2)$, the
wavefunctions are determined unambiguously by taking the limits at
$P$ approached from either sides along the line $K=1$. For instance,
when $K=1$, $J_R=2+\epsilon$, $\epsilon\rightarrow 0^+$, we have the
normalized state
\begin{eqnarray}
|\Psi_{(0^+)}\rangle_{+}&=&\frac{1}{2}[
-|\Uparrow\uparrow;\Downarrow\downarrow\rangle
-|\Downarrow\downarrow;\Uparrow\uparrow\rangle\nonumber\\
&&+|\Downarrow\uparrow;\Uparrow\downarrow\rangle
+|\Uparrow\downarrow;\Downarrow\uparrow\rangle].
\end{eqnarray}
While when $K=1$, $J_R=2-\epsilon$, $\epsilon\rightarrow 0^+$, we
have
\begin{eqnarray}
|\Psi_{(0^+)}\rangle_{-}&=&\frac{1}{2{\sqrt 3}}[
2|\Uparrow\downarrow;\Uparrow\downarrow\rangle
+2|\Downarrow\uparrow;\Downarrow\uparrow\rangle
-|\Uparrow\uparrow;\Downarrow\downarrow\rangle\nonumber\\
&&-|\Downarrow\downarrow;\Uparrow\uparrow\rangle
-|\Downarrow\uparrow;\Uparrow\downarrow\rangle
-|\Uparrow\downarrow;\Downarrow\uparrow\rangle].
\end{eqnarray}
Therefore, $_{+}\langle\Psi_{(0^+)}|\Psi_{(0^+)}\rangle_{-}=0$.
 This result demonstrates a discontinuity of the
groundstate wavefunction across the singular point.


\begin{thebibliography}{99}
\bibitem{note00}
There are two types of particle-hole symmetry in the TIKM with
lattice inversion symmetry, depending on the inter-impurity distance
being even or odd lattice spacing, respectively. Usually, only the
first type with even inter-impurity distance is the required
symmetry which may guarantee a critical point with groundstate
degeneracy\cite{Jones89,Affleck2}. As an effective spin-only model,
Eq.(\ref{h1}) does not show such distinction for even/odd
inter-impurity distance. Instead, the required groundstate
degeneracy can arise by tuning the cross Kondo coupling, see the
following discussions.
\bibitem{note0}
The  discontinuity in the first derivative of negativity usually
comes from the requirement of nonnegative concurrence\cite{Yang}.
Its relation with a true quantum phase transition should be further
corroborated by checking the nonanalyticity of groundstate energy or
groundstate wavefunction.
\bibitem{note1} In some circumstances such as in psudo-gaped or dissipative
hosts the SIKE measured by the von Neumann entropy can show a peak
feature at the quantum critical points\cite{Hur1,Hur2}.
\bibitem{note2} In a spin chain model with two
magnetic impurites several impurity-related entanglement quantities
measured by the negativity display detectable features in their
derivatives at the quantum phase transition \cite{Bayat}. These
quantities do not exhibit apparant variation of the Kondo effect
across the transition. More recently, the Schmidt gap has been
proposed as an order parameter measuring the impurity quantum phase
transitions\cite{Bayat14}.
\bibitem{note3}
In the conformal field theory and non-Abelian bosonization
approaches the enlarged hidden $SO(7)$ symmetry at the non-trivial
critical point is identified\cite{Affleck2,Gan}.
\end{thebibliography}
\end{document}